# An OAI-PMH-based Web Service for the Generation of Co-Author Networks

*Philipp Schaer, Thomas Lüke, Philipp Mayr, Peter Mutschke*

GESIS – Leibniz Institute for the Social Sciences,
Unter Sachsenhausen 6-8, 50667 Cologne, Germany

**Abstract**

We will present a new component of our technical framework that was built to provide a brought range of reusable web services for the enhancement of typical scientific retrieval processes. The proposed component computes betweenness of authors in co-authorship networks extracted from publicly available metadata that was harvested using OAI-PMH.

## Introduction

At ISI 2011 we presented a set of different retrieval-supporting techniques that can be used to surpass typical problems during the search process for scientific literature, such as the vagueness between search and indexing terms and the information overload by the amount of result records returned (Mayr et al. 2011). Among these was the so-called search term recommender which assists users during the query formulation phase by automatically suggesting appropriate thesaurus terms. Furthermore, alternative ranking methods were proposed that rely on the coreness of journals for a scientific field, detected by applying Bradford's law of scattering, and the centrality of authors, based on their betweenness in co-authorship networks. To enable the reusability of the services we developed the IRSA framework that implements all these methods as reusable web services to be used as internal or external services in a scientific retrieval environment. In 2012 we presented the first public version of IRSA that allows building custom search term recommenders based on previously harvested OAI-PMH metadata sets (Schaer et al. 2012).

In this poster and demo presentation we will show the next iteration of this development focussing on the online calculation of co-authorship networks and the extraction of central authors.

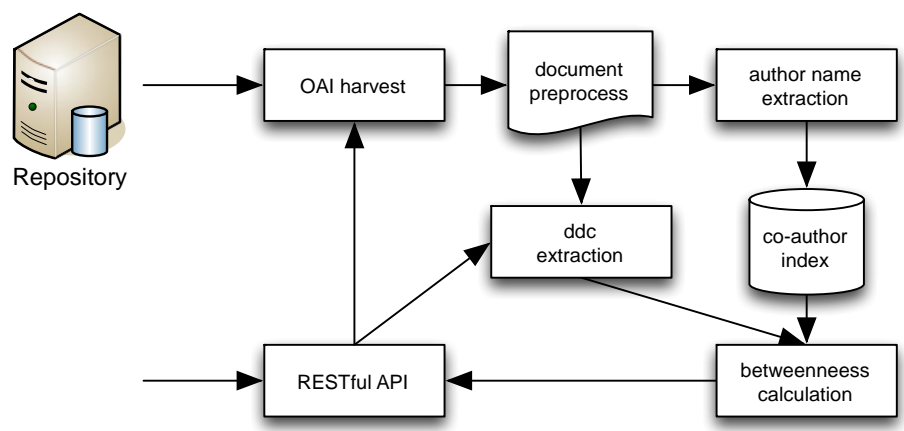

Figure 1: Overview on the workflow of the proposed software module.

## Workflow and Outcome

Figure 1 gives a rough overview on the workflow of the training and querying of the proposed web service. The system can be configured to harvest OAI metadata from public repositories (certified by DINI) from which the available DDC classification, author names and co-authorships are extracted and stored in a co-author index. The system can then be queried via a RESTful API and returns with the most central authors for each e.g. DDC classification group as XML formatted files or as PNG network plots. This way the centrality information can be used to gain an overview on the co-authorships in the specific repository and the DDC sub-categories.

## References

Mayr, P., Mutschke, P., Petras, V., Schaer, P. & Sure-Vetter, Y. (2011). Applying Science Models for Search. Proceedings des 12. Internationalen Symposiums für Informationswissenschaft (ISI 2011). Boizenburg: Verlag Werner Hülsbusch, pp. 184–196.

Schaer, P., Lüke, T. & Van Hoek, W. (2012). Building Custom Term Suggestion Web Services with OAI-Harvested Open Data. Proceedings of the 64. DGI Annual Meeting and 2nd DGI-Conference (DGI 2012), pp. 389–396.